\pdfoutput=1

\documentclass[twocolumn, trackchanges]{aastex61}
\received{2017 June 26}
\revised{2017 August 17}
\accepted{2017 August 19}
\published{2017 September 14}
\submitjournal{\apjl}

\shorttitle{Eclipse Spectroscopy of WASP-12b}
\shortauthors{Bell et al.}

\usepackage{graphicx}
\usepackage{amsmath}
\usepackage{amssymb}
\usepackage{bm}
\usepackage{cleveref}
\newcommand{\RNum}[1]{\uppercase\expandafter{\romannumeral #1\relax}}

\begin{document}

\title{The Very Low Albedo of WASP-12b from Spectral Eclipse Observations with \textit{Hubble}}

\correspondingauthor{Taylor J. Bell}
\email{taylor.bell@mail.mcgill.ca}

\author[0000-0003-4177-2149]{Taylor J. Bell}
\altaffiliation{McGill Space Institute; Institute for Research on Exoplanets}
\affil{Department of Physics, McGill University, 3600 rue University, Montr\'eal, QC H3A 2T8, Canada}

\author[0000-0002-6500-3574]{Nikolay Nikolov}
\affil{Department of Physics and Astronomy, University of Exeter, Exeter EX4 4QL, UK}

\author[0000-0001-6129-5699]{Nicolas B. Cowan}
\altaffiliation{McGill Space Institute; Institute for Research on Exoplanets}
\affil{Department of Physics, McGill University, 3600 rue University, Montr\'eal, QC H3A 2T8, Canada}
\affil{Department of Earth \& Planetary Sciences, McGill University, 3450 rue University, Montr\'eal, QC H3A 0E8, Canada}

\author[0000-0003-3726-5419]{Joanna K. Barstow}
\affil{Department of Physics and Astronomy, University College London, Gower Street, London WC1E 6BT, UK}

\author[0000-0002-7129-3002]{Travis S. Barman}
\affil{Lunar and Planetary Laboratory, University of Arizona, 1629 E. University Boulevard, Tucson, AZ 85721, USA}

\author{Ian J. M. Crossfield}
\affil{Department of Physics, Massachusetts Institute of Technology, Cambridge, MA, USA}

\author[0000-0002-9308-2353]{Neale P. Gibson}
\affil{Astrophysics Research Centre, School of Mathematics and Physics, Queens University Belfast, Belfast BT7 1NN, UK}

\author{Thomas M. Evans}
\affil{Department of Physics and Astronomy, University of Exeter, Exeter EX4 4QL, UK}

\author[0000-0001-6050-7645]{David K. Sing}
\affil{Department of Physics and Astronomy, University of Exeter, Exeter EX4 4QL, UK}

\author{Heather A. Knutson}
\affil{Division of Geological and Planetary Sciences, California Institute of Technology, Pasadena, CA 91125, USA}

\author[0000-0003-3759-9080]{Tiffany Kataria}
\affil{Jet Propulsion Laboratory, California Institute of Technology, 4800 Oak Grove Drive, Pasadena, CA 91109, USA}

\author[0000-0003-3667-8633]{Joshua D. Lothringer}
\affil{Lunar and Planetary Laboratory, University of Arizona, 1629 E. University Boulevard, Tucson, AZ 85721, USA}

\author[0000-0001-5578-1498]{Bj\"orn Benneke}
\affil{Department of Physics, Universit\'e de Montr\'eal, 2900 boul.\ \'Edouard-Montpetit, Montr\'eal, QC H3T 1J4, Canada}

\author[0000-0001-5232-9957]{Joel C. Schwartz}
\altaffiliation{McGill Space Institute; Institute for Research on Exoplanets}
\affil{Department of Physics, McGill University, 3600 rue University, Montr\'eal, QC H3A 2T8, Canada}
\affil{Department of Earth \& Planetary Sciences, McGill University, 3450 rue University, Montr\'eal, QC H3A 0E8, Canada}




\begin{abstract}
	We present an optical eclipse observation of the hot Jupiter WASP-12b using the Space Telescope Imaging Spectrograph on board the Hubble Space Telescope. These spectra allow us to place an upper limit of $A_g < 0.064$ (97.5\% confidence level) on the planet's white light geometric albedo across 290--570 nm. Using six wavelength bins across the same wavelength range also produces stringent limits on the geometric albedo for all bins. However, our uncertainties in eclipse depth are $\sim$40\% greater than the Poisson limit and may be limited by the intrinsic variability of the Sun-like host star --- the solar luminosity is known to vary at the $10^{-4}$ level on a timescale of minutes. We use our eclipse depth limits to test two previously suggested atmospheric models for this planet: Mie scattering from an aluminum-oxide haze or cloud-free Rayleigh scattering. Our stringent nondetection rules out both models and is consistent with thermal emission plus weak Rayleigh scattering from atomic hydrogen and helium. Our results are in stark contrast with those for the much cooler HD 189733b, the only other hot Jupiter with spectrally resolved reflected light observations; those data showed an increase in albedo with decreasing wavelength. The fact that the first two exoplanets with optical albedo spectra exhibit significant differences demonstrates the importance of spectrally resolved reflected light observations and highlights the great diversity among hot Jupiters.
\end{abstract}
\keywords{planets and satellites: atmospheres --- stars: individual (WASP-12) --- techniques: photometric}



\section{Introduction}\label{sec:intro}
Thermal measurements of hot Jupiters suggest that these gas giant exoplanets often have moderate Bond albedos ($A_B \approx 0.4$, the fraction of incident energy reflected to space; \citealt{schwartz2017}). However, many previous searches for reflected light from hot Jupiters have found little-to-none at optical wavelengths where the host star emits most of its energy \citep[geometric albedo $A_g < 0.1$; e.g.,][]{rowe2008, kipping2010, heng2013, dai2017}. It is unclear what is causing this apparent contradiction between constraints from thermal emission and optical reflection. Previous Hubble Space Telescope (\textit{HST}) eclipse observations of HD~189733b with the Space Telescope Imaging Spectrograph (STIS) showed an increase in reflectivity toward bluer wavelengths which may, at least in part, explain the discrepancies between these two techniques \citep{evans2013}.

A direct way to probe the back scattering efficiency of a hot Jupiter's atmosphere is observing the planet at optical wavelengths (where thermal emission is negligible) during eclipse, when the planet is near full phase and passes behind its host star. This method requires at least an order of magnitude higher photometric precision than transit observations of the same planet because the planet will be fainter than its host star, while the occulted area remains the same.

Observing an atmosphere at different orbital phases can provide further information about the scattering particles \citep[e.g.,][]{demory2013, esteves2013, heng2013, munoz2015, shporer2015, oreshenko2016}. \citet{parmentier2016} suggested a connection between reflected light phase curve measurements and a sequence of condensate cloud models, but this only covered temperatures up to $T_{\rm eq}\sim2200$~K: well below the equilibrium temperature of WASP-12b ($T_{\rm eq}=2580$~K; \citealt{collins2017}).

WASP-12b orbits a G0V star with an orbital period of 1.09 days \citep{hebb2009}. While the host star is fairly faint (V~=~12), \mbox{WASP-12b's} close semi-major axis and large radius ($a$~=~0.0234~au, $R_p=1.90~R_J$, \mbox{$R_{p}=0.19~R_{*}$}; \citealt{collins2017}) make it an excellent target for detailed study. Transit observations of WASP-12b range from 0.3 to 4.5~$\mu$m, and eclipse observations range from 0.9 to 8.0~$\mu$m \citep[e.g.,][]{hebb2009, lopezMorales2010, campo2011, madhusudhan2011, cowan2012, crossfield2012, copperwheat2013, fohring2013, sing2013, swain2013, stevenson2014a, stevenson2014b, croll2015, sing2016}. This work presents the first optical eclipse measurement of WASP-12b.

The atmospheric composition of WASP-12b has been extensively studied \citep[e.g.,][]{madhusudhan2011, crossfield2012, swain2013, stevenson2014b}, with initial claims of a C/O ratio greater than unity. This was first challenged by \citet{crossfield2012} and \citet{cowan2012}, who instead reported an isothermal photosphere for WASP-12b. The recent detection of water in the planet's atmosphere has now firmly refuted the carbon-rich hypothesis \citep{kreidberg2015}. \citet{sing2013} found that the best-fit model for \mbox{WASP-12b} transmission spectroscopy was Mie scattering by an aluminum-oxide (Al$_2$O$_3$) haze. \citet{barstow2016} found that an optically thick Rayleigh scattering aerosol with a 0.01 mbar top pressure best described the transmission observations, but the model poorly described the steep increase in transit depth at optical wavelengths. \citet{schwartz2017} used thermal phase variations and eclipse depths to determine a Bond albedo of $A_B=0.2^{+0.1}_{-0.12}$ and a dayside effective temperature of $T_{\rm day}=\nolinebreak2864\pm15$~K.

\section{Observations and Data Reduction}\label{sec:observations}
On 2016 October 19, a single eclipse of WASP-12b was observed with five \textit{HST} orbits, using the STIS G430L grating (290--570 nm). The first \textit{HST} orbit has significantly worse systematics than the four later orbits as a result of the repointing of the telescope, so these data were removed from the subsequent analysis. This left two \textit{HST} orbits out of eclipse (one before and one after) when the planet and host star were both visible with the planet near full phase, as well as two \textit{HST} orbits during eclipse when the planet was behind its host star, leaving only the star's light visible. These observations were granted as a part of programme GO-14797 (PI: Crossfield).

We used the same data collection method as previously used for similar observations \citep{sing2011, sing2013, sing2016, evans2013}. The subarray readout mode with a wide $52\arcsec\times2\arcsec$ slit was used to minimize time-varying slit losses; this produced $1024 \times 128$ pixel images. In previous \textit{HST}/STIS observations, the first frame from each \textit{HST} orbit had systematically lower counts, so a 1~s dummy exposure was obtained at the beginning of each orbit, which successfully mitigated this systematic effect. This dummy exposure was then followed by 10 science exposures lasting 279~s each (the maximum recommended duration to avoid excessive cosmic-ray hits). Our final, analyzed dataset thus contains 40 exposures collected over 331 minutes.

The raw STIS data were reduced (bias-, dark-, and flat-corrected) using the latest version of the \texttt{CALSTIS1} pipeline and the relevant up-to-date calibration frames. Cosmic-ray events were identified and removed following \citet{nikolov2014}, as were all pixels identified as ``bad'' by \texttt{CALSTIS}. Overall, $\sim$9\% of the pixels in each 2D spectrum were affected by cosmic-rays with another $\sim$5\% identified as ``bad'', resulting in a total of $\sim$14\% interpolated pixels.

Next, the IRAF procedure \texttt{apall} was used to extract spectra from the calibrated .flt science files. We tested apertures between 9.0 and 17.0 pixels in intervals of 2 pixels and found that an 11.0 pixel aperture resulted in the lowest lightcurve residual scatter after fitting the white light data. However, the difference between apertures was minute ($\sim$1~ppm). We then used cross-correlation to correct for subpixel shifts along the dispersion axis. The x1d files from \texttt{CALSTIS} were then used to calibrate the wavelength axis. Finally, both ``white light'' and six spectral channel lightcurves were produced by integrating the appropriate flux from each bandpass.

WASP-12b's host star WASP-12A is also orbited by two M-dwarf companions bound in a binary system 1.06$\arcsec$ away from WASP-12A \citep{bergfors2011, sing2013, bechter2014}. For our observations, the spectrograph slit orientation was chosen to be perpendicular to the line connecting WASP-12A and WASP-12(B,C) to allow maximal separation in the spatial direction of the resulting FITS files. The spectrum of the stellar companions is visually distinguishable from WASP-12A in the raw spectra and does not fall within our small spatial-axis aperture.

\section{Lightcurve Analysis}\label{sec:analysis}
The top panel of \Cref{fig:lightcurve} shows the raw lightcurve binned across the entire STIS G430L bandpass (``white light''). There is a strong, repeated trend in flux, with exposures from each orbit appearing to follow a roughly polynomial trend. This systematic is well known and is believed to be the result of the thermal cycle of \textit{HST} throughout its orbit as well as the movement of the spectral trace on the detector \citep[e.g.,][]{brown2001, sing2011, huitson2012, evans2013}.

These systematic trends are also observed during \textit{HST}/STIS observations of planetary transits, and a standard approach to remove them is assuming polynomial variations as a function of auxiliary variables \citep[e.g.,][]{sing2011, huitson2012}. More recently, \citet{gibson2011} used Gaussian processes (GPs) to model the \textit{HST} systematics, as the choice of polynomial model can potentially bias the results. For this reason, we modelled the systematics with the GP library \texttt{george} \citep{foremanMackey2015_george} and used the same method as \citet{evans2013}. A detailed discussion of modelling systematics with GPs can be found in \citet{gibson2012a, gibson2012b, gibson2013}. We also attempted to fit the systematic variations with a polynomial model, which gave results consistent with our GP model.

\subsection{Gaussian Process Model}\label{subsec:GP}

The likelihood of a GP model is described as a multivariate normal distribution with
\begin{equation}
	p(\boldsymbol{f}|\boldsymbol{X}, \boldsymbol{\theta}, \boldsymbol{\Omega}, \boldsymbol{t}) = \mathcal{N}(\boldsymbol{E}(\boldsymbol{\Omega}, \boldsymbol{t}), \boldsymbol{\Sigma}(\boldsymbol{X}, \boldsymbol{\theta})) \,,
\end{equation}
where $\boldsymbol{f}$ is the 40 measured fluxes, $\boldsymbol{E}$ is the eclipse function, and $\boldsymbol{\Sigma}$ is the \textit{kernel} (covariance matrix). The time at the midpoint of each exposure is represented by $\boldsymbol{t}$. Further, $\boldsymbol{X}=[\boldsymbol{\phi}, \boldsymbol{\psi}]^T$ is the matrix of covariates, where $\boldsymbol{\phi}$ is the orbital phase of \textit{HST}, and $\boldsymbol{\psi}$ is the slope of the spectral trace on the detector (computed using IRAF's \texttt{apall} procedure). These two covariates were selected as they provided the lowest scatter in the residuals after calibration. We also tested the inclusion of two additional covariates: the y-intercept of the spectral trace on the detector and the measured shifts of the spectral trace along the dispersion axis. However, the inclusion of these additional covariates did not significantly impact our results or uncertainties, likely because the covariates themselves are significantly correlated with the other covariates.

Our eclipse parameters are given by $\boldsymbol{\Omega}$~=~$[\alpha, \delta, \beta]^T$, where $\alpha$ is the baseline flux consisting of light emitted from both the planet and star, $\delta$ is the fractional eclipse depth ($\delta = F_{\text{planet}}/F_{\text{star}}$), and $\beta$ describes a constant rate of change in the baseline flux over time. Since we did not observe during eclipse ingress or egress, we used a boxcar function to describe the eclipse signal, with
\begin{equation}
\begin{aligned}
	&E_i = \alpha(1 - \delta B_i)(1 + \beta (t_i-t_0)) \\
	&B_i = 
		\begin{cases} 
    		0 & i \in \text{Orbit 2 or 5} \\ 1 & i \in \text{Orbit 3 or 4}\,,
		\end{cases}
\end{aligned}
\end{equation}%
where $t_0$ is the time of the first exposure.

Our GP parameters are given by $\boldsymbol{\theta}$~=~$[C, L_\phi, L_\psi, \sigma_w]^T$, where $C^2$ is the maximum covariance, $L_\phi$ and $L_\psi$ are covariance lengthscales, and $\sigma_w$ is the white noise level. We adopted the squared-exponential kernel:
\begin{equation}
	\boldsymbol{\Sigma}_{nm} = 
		C^2 \exp \Bigg[-\sum^{1}_{i=0}\frac{(\boldsymbol{X}_{in}-\boldsymbol{X}_{im})^2}{L_{i}^2}\Bigg] 
		+ \delta_{nm}\sigma_w^2 \,,
\end{equation}
where $L_i=[L_\phi, L_\psi]_i$ and $\delta_{nm}$ is the Kronecker delta function. This kernel can be simply understood as requiring that observations be strongly correlated if they have similar spectral trace slope and \textit{HST} orbital phase, while observations further from each other in covariate space are more weakly correlated. This then describes a smoothly varying function of the covariates, with the addition of white noise.

The final model is then given by
\begin{equation}
	f^* = \boldsymbol{\mu}(\boldsymbol{\phi}, \boldsymbol{\psi})+\boldsymbol{E}(\boldsymbol{\Psi}) \,,
\end{equation}
where $\boldsymbol{\mu}(\boldsymbol{\phi}, \boldsymbol{\psi})$ is the GP model mean.

The Markov Chain Monte Carlo (MCMC) ensemble sampler software \texttt{emcee} \citep{foremanMackey2015_emcee} was used to explore the seven parameters, determining the most likely parameter values and their uncertainties. For computational reasons, the variables used in this MCMC were \{$\delta$, $\ln(\alpha)$, $\beta$, $\ln(L_\phi)$, $\ln(L_\psi)$, $\ln(\sigma_w^2)$, and $\ln(C^2)$\}. Using logarithms removes the need to use a prior to obtain strictly positive values. While the eclipse depth, $\delta$, should be strictly positive, we allowed for negative values to ensure an unbiased estimate. A uniform prior was used so $\ln(L_\phi) < 0$ and $\ln(L_\psi) < 0$, which has the effect of ensuring that these lengthscales are within a few orders of magnitude of the variations in the covariates. For un-normalized white light data, the best-fit values from a $10^6$ step MCMC chain were \{\mbox{$\delta=(-5.3\pm7.4)\times\nobreak10^{-5}$,} \mbox{$\alpha=(4.198^{+0.012}_{-0.008})\times10^7$}, \mbox{$\beta=0.0056\pm0.0015$}, \mbox{$L_\phi=0.05^{+0.19}_{-0.03}$,} \mbox{$L_\psi=0.03^{+0.27}_{-0.03}$,} \mbox{$\sigma_w^2=(5.2^{+1.7}_{-1.3})\times10^7$}, and \mbox{$C^2=(8^{+14}_{-5})\times10^7$}\}.

\begin{figure}
	\centering
	\includegraphics[width=\linewidth]{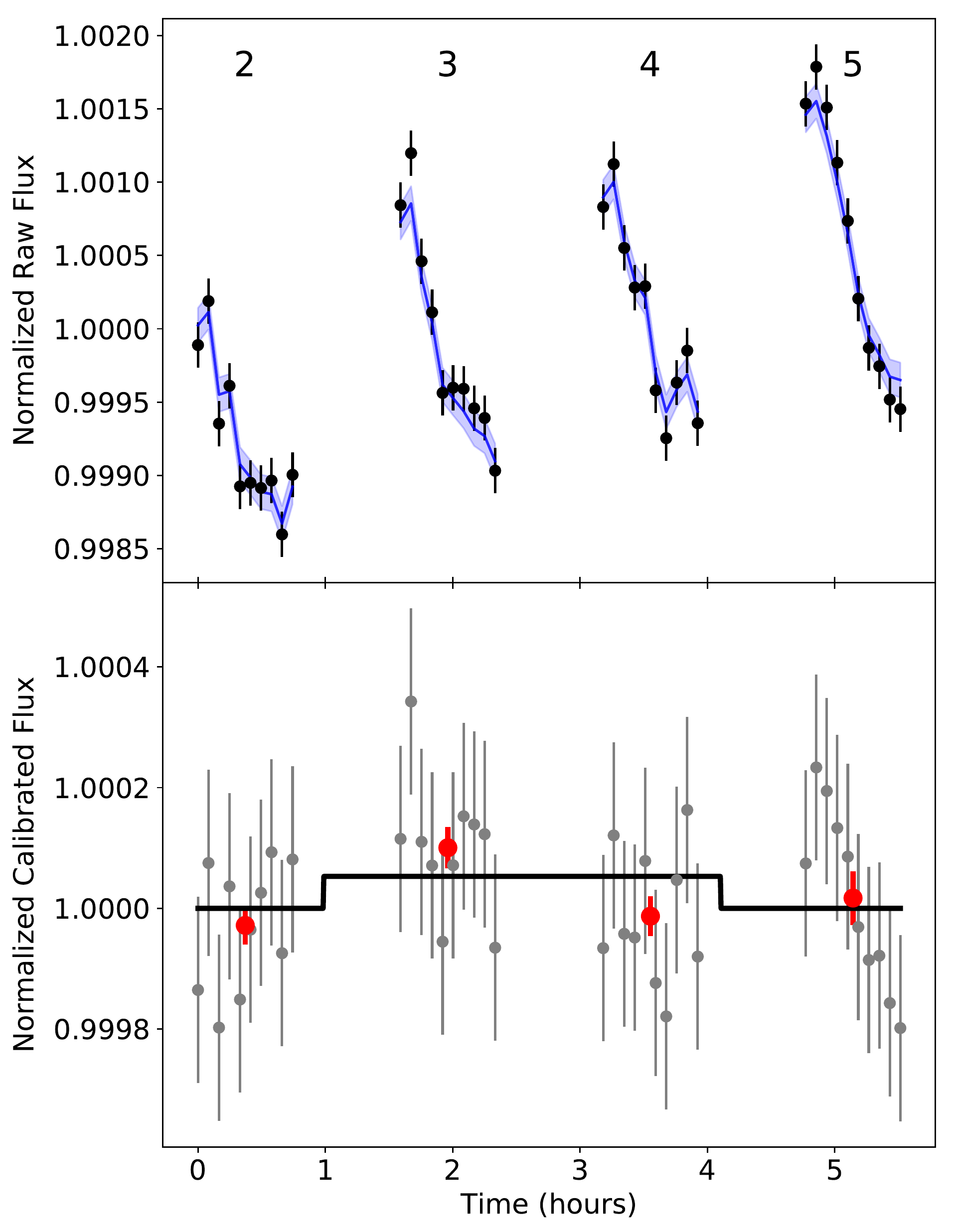}
	\caption{\textit{Top:} raw flux with the entire spectral range of \textit{HST}/STIS binned into a single white lightcurve. The median systematic model and $1\sigma$ model uncertainty are shown with a blue line and blue shaded region, respectively. Each individual \textit{HST} orbit is labelled. \textit{Bottom:} the white light data after calibration using a Gaussian Process are shown in grey. Also shown in red are the binned fluxes for each \textit{HST} orbit, although these were not used during fitting. Overplotted is the best-fit eclipse signal that corresponds to a wavelength-averaged geometric albedo of $A_g=-0.035$ ($A_g < 0.064$ at 97.5\% confidence). All plotted error bars in both panels only capture uncorrelated, white noise.}\label{fig:lightcurve}
\end{figure}

The top panel of \Cref{fig:lightcurve} shows the median model and uncertainty from a $10^6$ step MCMC chain overplotted on the raw white light flux measurements. The bottom panel of \Cref{fig:lightcurve} shows the lightcurve produced by dividing the raw spectra by the median model (excluding the change in flux during eclipse), with the median eclipse model overplotted. The clear linear trend in the calibrated flux of \textit{HST} orbit \#5 (bottom panel of \Cref{fig:lightcurve}) shows that there is still substantial correlated noise in the data that could not be described by any of the four considered covariates.

\begin{figure*}
	\centering
	\includegraphics[width=0.80\linewidth]{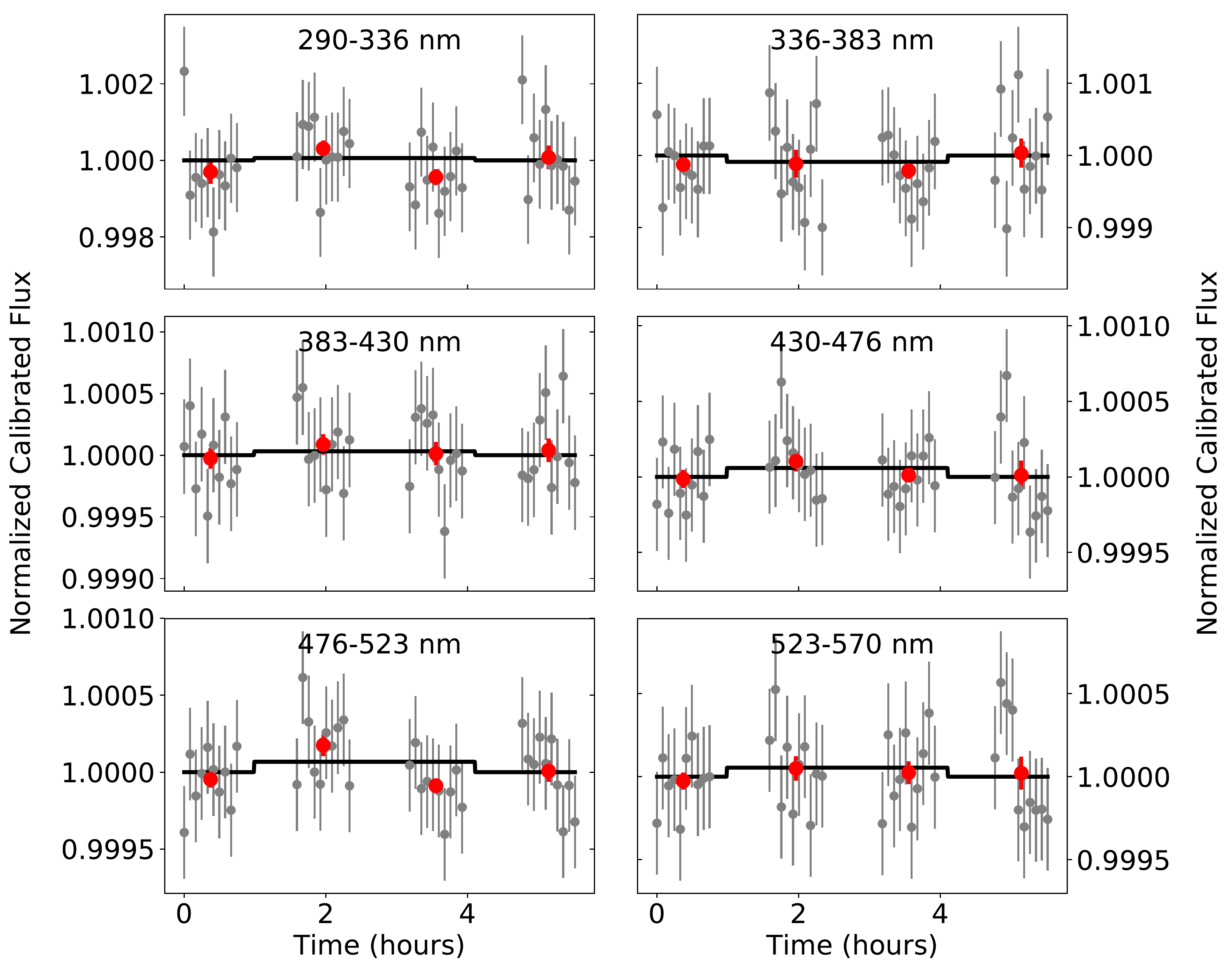}
	\caption{Lightcurves for each spectral channel after calibration using a Gaussian process are shown in grey, with the best-fit eclipse signal overplotted. Also shown in red are the binned fluxes for each \textit{HST} orbit, although these were not used during fitting. All plotted error bars only capture uncorrelated, white noise; where no error bar is visible, it is smaller than the point size used.}\label{fig:colourCurves}
\end{figure*}

\section{Results}\label{sec:results}
The STIS G430L spectra were binned into six spectral channels to allow moderate wavelength resolution while keeping uncertainties on each channel sufficiently small to be able to test atmospheric models. Each spectral channel was modelled independently using the GP method described above. Lightcurves after GP calibration are shown for each spectral channel in \Cref{fig:colourCurves}, and the relevant results are tabulated in Table \ref{tab:depths}. Eclipse depths were found using the median value from a $10^6$ step MCMC chain, while the 84 and 97.5 percentiles were used to determine upper limits. The larger uncertainties in eclipse depth at shorter wavelengths are due to lower stellar flux and detector sensitivity.

Because WASP-12b is so strongly irradiated, the peak of its thermal emission is expected to be at $\sim$1~$\mu$m for a $\sim$2800~K dayside temperature \citep{schwartz2017}. For this reason, we calculated the predicted eclipse depths due to thermal radiation from WASP-12b, assuming a $T=3000$~K blackbody for WASP-12b (hotter than inferred from infrared observations due to the greater depth of the optical photosphere; \citealt{cowan2011}) and a standard G0V spectrum from \citet{pickles1998} for WASP-12. These depths ($\delta_{\rm thermal}$) are summarized in Table \ref{tab:depths} and are all within our 97.5\% confidence interval upper limits.

If interpreted as solely due to reflected light, eclipse depths can be converted to geometric albedo using
\begin{equation}
	A_g = \delta \Bigg(\frac{R_p}{a}\Bigg)^{-2} \,,
	\label{eqn:geometricAlbedo}
\end{equation}
where $R_p=1.90$~$R_J$ is the radius of the planet, and $a=0.0234$~AU is its orbital semi-major axis \citep{collins2017}. Applying \Cref{eqn:geometricAlbedo} to the best-fit eclipse depths and their corresponding upper limits gives constraints on the geometric albedo across the STIS G430L wavelength range (summarized in Table \ref{tab:depths})

\begin{deluxetable*}{cccccc}
	\tablecaption{Eclipse Depths and Geometric Albedos}
	\tablehead{\colhead{Wavelengths} & \twocolhead{Eclipse Depth, $\delta$ (ppm)} & \colhead{$\delta_{\rm thermal}$} & \twocolhead{Geometric Albedo, $A_g$} \\ 
	\colhead{(nm)} & \colhead{Best fit} & \colhead{97.5\% Upper Limit} & \colhead{(ppm)} & \colhead{Best Fit} & \colhead{97.5\% Upper Limit} } 
	\tablenum{1}\label{tab:depths}
	\startdata
		290~--~570 & $-53\pm74$ & 96 & 56 & $-0.035\pm0.050$ & 0.064 \\ \hline
		290~--~336 & $-60\pm540$ & 1020 & 10 & $-0.04\pm0.36$ & 0.68 \\
		336~--~383 & $~~90\pm290$ & 670 & 20 & $~~0.06\pm0.20$ & 0.45 \\
		383~--~430 & $-30\pm180$ & 330 & 40 & $-0.02\pm0.12$ & 0.22 \\
		430~--~476 & $-60\pm130$ & 210 & 60 & $-0.039\pm0.089$ & 0.14 \\
		476~--~523 & $-70\pm130$ & 190 & 100 & $-0.045\pm0.087$ & 0.13 \\
		523~--~570 & $-50\pm150$ & 240 & 160 & $-0.036\pm0.098$ & 0.16 \\
	\enddata
\end{deluxetable*}

Our reported uncertainties on eclipse depths are $\sim$40\% higher than the photon limit. The increased scatter in our data may be the result of incomplete modelling of the systematic noise. Alternatively, our uncertainties may be limited by intrinsic stellar variability. Given the slow rotation period of WASP-12 compared to the observing window \citep[$P_{rot}\gtrsim23$~days given $v\sin i < 2.2$~km/s;][]{hebb2009}, variability due to stellar rotation (e.g. starspots passing in and out of view) should not significantly affect our observations. However, our Sun's total irradiance (spatially and spectrally integrated) is known to vary at the $10^{-4}$ level on timescales of minutes to hours as a result of solar convection and oscillations \citep{kopp2016}. Given the G0V spectral class of WASP-12, similar variations may also be present and may explain the greater than Poisson limit uncertainties as well as the residual correlated noise in the calibrated time-series spectra.

\section{Discussion and Conclusions}\label{sec:conclusion}
We use the \texttt{NEMESIS} spectral retrieval tool \citep{irwin2008, barstow2014} to produce predicted model spectra given two previously proposed models for WASP-12b: an Al$_2$O$_3$ haze and a cloud-free atmosphere. \texttt{NEMESIS} is not a radiative equilibrium code; rather, it takes an atmospheric model and calculates incident and emergent flux and will not take into account heating from incoming stellar radiation. The limits from our \textit{HST}/STIS eclipse observations firmly reject both models; we find $\chi^2$ per datum ($\chi^2/N_{obs}$, \mbox{$N_{obs}=6$}) of 41 and 10 for the Al$_2$O$_3$ haze and cloud-free models.

\begin{figure}
	\centering
	\includegraphics[width=\linewidth]{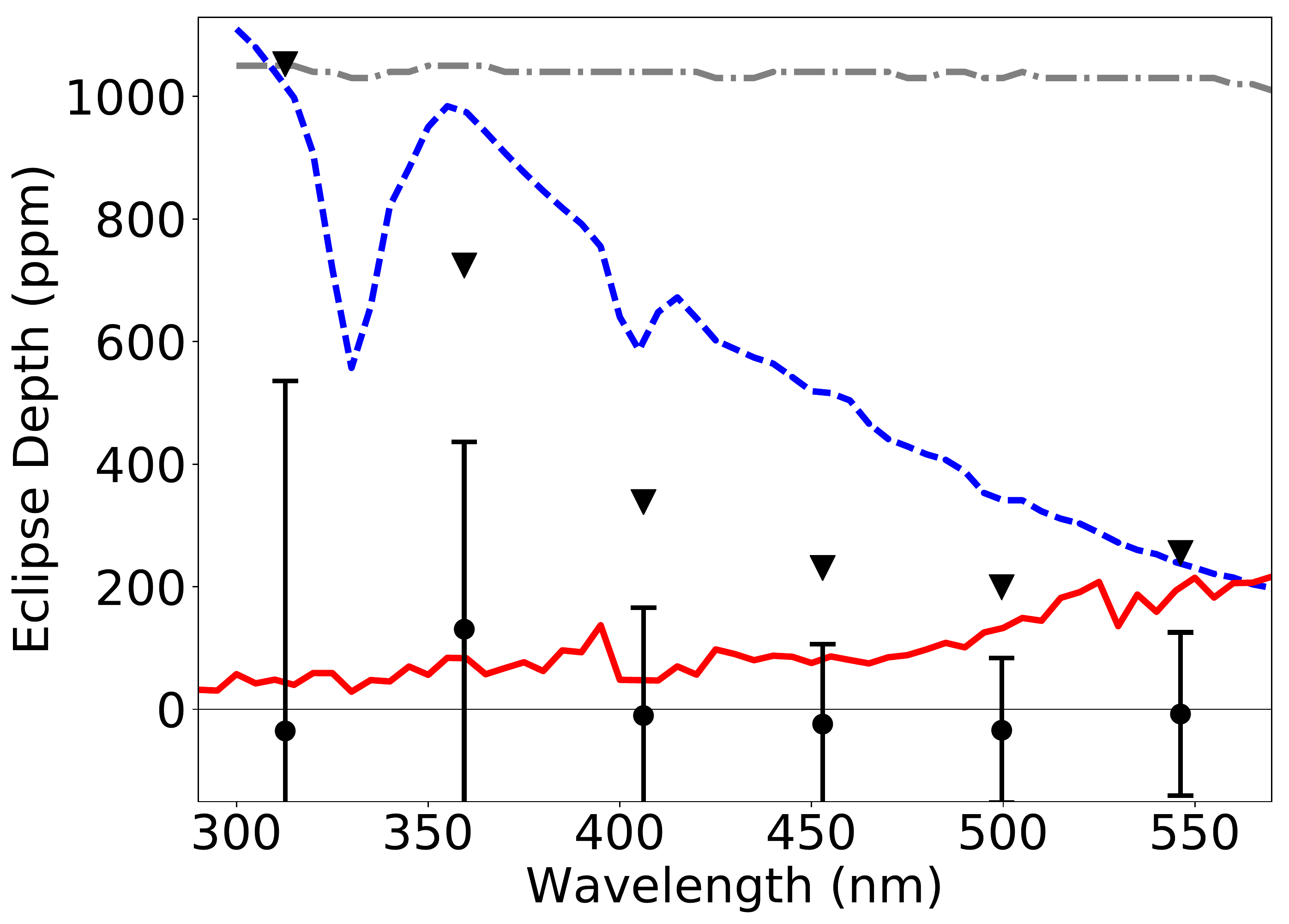}
	\caption{Best-fit eclipse depths and $1\sigma$ uncertainties are shown with black points and error bars, with black triangles denoting 97.5\% confidence upper limits. Previously proposed models for WASP-12b made with \texttt{NEMESIS} are shown with a grey, dashed-dotted line (aluminum-oxide haze) and a blue, dashed line (cloud-free). The \textit{HST}/STIS data firmly reject both models and are instead consistent with the thermally dominated \texttt{PHOENIX} model shown with a red solid line.}\label{fig:spectra}
\end{figure}

\begin{figure}
	\centering
	\includegraphics[width=\linewidth]{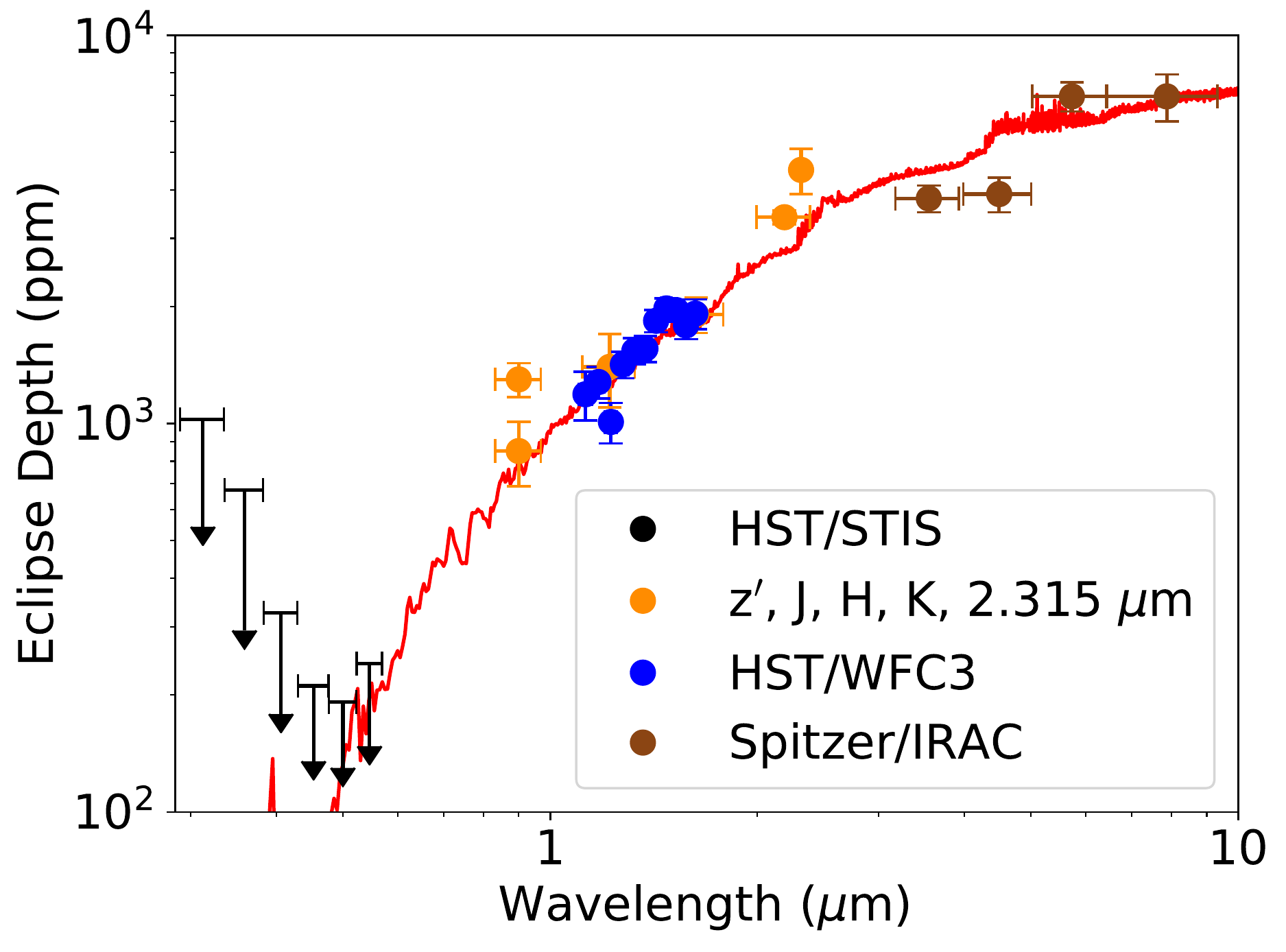}
	\caption{Our \textit{HST}/STIS 97.5\% confidence interval upper limits on the eclipse depth for each of the six considered spectral channels are shown with black arrows. All eclipse depths aside from \textit{HST}/STIS are taken from \citet[and references therein]{stevenson2014b}. The red line is the same as in \Cref{fig:spectra}.}\label{fig:allObs}
\end{figure}

Given its exceedingly high equilibrium temperature ($T_{\rm eq}=2580$~K; \citealt{collins2017}), WASP-12b would technically lie within \citeauthor{sudarsky2000}'s (\citeyear{sudarsky2000}) Class V ($T_{\rm eff}>1500$~K) but is far hotter than any planet they considered. On the planet's dayside, WASP-12b is far too hot for condensates to form \citep{wakeford2017}. However, temperatures near the planet's day--night terminator, and across the planet's nightside, may be cool enough to allow for the formation of condensates that could affect transmission spectroscopy without significantly affecting dayside eclipse spectroscopy.

Also, it is expected that Na~\RNum{1} absorption (which is important at lower temperatures) will not contribute much to the low albedo of WASP-12b as most of the sodium will be ionized on the hot dayside. Instead, it is expected that the atmosphere will be dominated by Rayleigh scattering from atomic hydrogen and helium, with a small contribution from electron scattering. The red line in \Cref{fig:spectra,fig:allObs} shows the predicted eclipse depth (binned to a resolution of 1 point per 5~nm) from \citet{crossfield2012} made with the \texttt{PHOENIX} atmosphere code adapted for hot Jupiters as described in \citet{barman2001, barman2005}. In this model, reflected light makes up $\lesssim10$\% of the eclipse depth ($A_g \lesssim 0.002$) at the shortest wavelengths and $\ll 1$\% of the eclipse depth at infrared wavelengths; the remainder of the eclipse depth is due to thermal emission. This model gives a $\chi^2$ per datum of 0.9 for our \textit{HST}/STIS data ($N_{obs}=6$), but a worse $\chi^2$ per datum of 3 for all of the data plotted on \Cref{fig:allObs} ($N_{obs}=21$).

There are significant differences between the \texttt{PHOENIX} model and the cloud-free model produced by \texttt{NEMISIS}, including but not limited to the inclusion of atomic hydrogen opacities (lines and bound-free opacities), as well as the typical opacities more commonly associated with cool stellar photospheres.  Also, the \texttt{PHOENIX} model results from a self-consistent calculation of the thermal structure, chemistry, line-by-line opacities (as well as scattering), and irradiation, thereby accounting for important changes that occur in the hot upper layers of WASP-12b (for example, the transition from H$_2$ to H at low pressures and the thermal ionization of Na and K).

Our observations cover the blackbody peak of \mbox{WASP-12} ($\sim$450~nm) and show that little of the incident radiation at these wavelengths is reflected by the planet. Geometric albedo is related to spherical albedo through a phase integral $q$ such that $A_s=qA_g$, and Bond albedo is equal to the flux-weighted, wavelength-averaged spherical albedo. If we assume diffuse scattering ($q=1.5$), our ``white light'' 97.5\% confidence upper limit on the geometric albedo across the STIS bandpass ($A_g<0.064$) suggests $<10\%$ of the energy received at these wavelengths is reflected. However, since the wavelengths observed cover only 36\% of the incident stellar energy, the Bond albedo is not well constrained by these measurements and is consistent with \citeauthor{schwartz2017}'s (\citeyear{schwartz2017}) measurement of $A_B = 0.2^{+0.1}_{-0.12}$.

Our results are in stark contrast with those for the much cooler HD 189733b, the only other hot Jupiter with spectrally resolved reflected light observations \citep{evans2013}; those data showed an increase in albedo with decreasing wavelength. The fact that the first two exoplanets with optical albedo spectra exhibit significant differences demonstrates the importance of spectrally resolved reflected light observations and highlights the great diversity among hot Jupiters.

\acknowledgments

T.J.B.\ acknowledges support from the McGill Space Institute Graduate Fellowship and from the FRQNT through the Centre de recherche en astrophysique du Qu\'ebec. J.K.B.\ acknowledges support from the Royal Astronomical Society Research Fellowship. I.J.M.C.\ was supported under contract with the Jet Propulsion Laboratory (JPL) funded by NASA through the Sagan Fellowship Program executed by the NASA Exoplanet Science Institute. The research leading to these results has received funding from the European Research Council under the European Union's Seventh Framework Programme (FP7/2007-2013) / ERC grant agreement No.~336792. This work is based on observations made with the NASA/ESA Hubble Space Telescope, obtained at the Space Telescope Science Institute, which is operated by the Association of Universities for Research in Astronomy, Inc., under NASA contract NAS 5-26555. The Al$_2$O$_3$ and cloud-free models tested in this work were made with the \texttt{NEMESIS} code developed by Patrick Irwin. We have also made use of free and open-source software provided by the Matplotlib, Python, and SciPy communities.

%

\vspace{5mm}
\facilities{HST(STIS)}

\end{document}